\documentclass[preprint,aps,prd,nofootinbib,superscriptaddress,11pt]{revtex4}
\usepackage{txfonts}
\usepackage[colorlinks, linkcolor=blue, citecolor=blue, urlcolor=blue]{hyperref}
\usepackage{float}
\usepackage{graphicx}
\usepackage{dcolumn}
\usepackage{bm}
\usepackage{amssymb}
\usepackage{latexsym}

\newcommand{\be}{\begin{equation}}
\newcommand{\ee}{\end{equation}}
\newcommand{\bq}{\begin{eqnarray}}
\newcommand{\eq}{\end{eqnarray}}

\begin{document}

\title{ A Possible Hermitian  Neutrino Mixing  Ansatz }

\author{ Wu-zhong Guo}\email{wuzhong@itp.ac.cn}
\author{Miao Li}\email{mli@itp.ac.cn}

\affiliation{Institute of Theoretical Physics, Chinese Academy of Sciences, Beijing 100190, China}
\affiliation{Kavli Institute for Theoretical Physics China, Chinese Academy of Sciences, Beijing 100190, China}
\affiliation{National Key Laboratory of Frontiers in Theoretical Physics, Chinese Academy of Sciences, Beijing 100190, China}

\begin{abstract}

Using a recent global analysis result after the precise measurement of $\theta_ {13}$, a possible Herimtian neutrino mixing ansatz is discussed, the mixing
matrix is symmetric and also symmetric with respect with the second diagonal line in the leading order. This leading order ansatz predicts
$\theta_{13}=12.2^\circ$. Next, consider the hierarchy structure of the lepton mass matrix as the origin of perturbation of the mixing matrix, we find that this ansatz with perturbation can fit current data very well.
\end{abstract}
\maketitle

\section{Introduction}
Many neutrino experiments have revealed that the three known light neutrinos must have finite but small masses and that different flavor neutrinos oscillate from one to another. The massive neutrino states $\nu_{L}$ are relateded to the flavor states $\nu_{fL}$ by the $3 \times 3$ PMNS matrix:
\begin{eqnarray}
\left(\begin{array}{c} \nu_e\\ \nu_{\mu} \\ \nu_{\tau}\end{array}
\right)_{L}=~ U_{PMNS}\left(\begin{array}{c} \nu_1 \\ \nu_2 \\ \nu_3
\end{array} \right)_{L}\end{eqnarray}
The PMNS matrix is unitary, and can be parameterized in the standard way, with three mixing angles and three CP-related phases:
\begin{eqnarray}
U_{PMNS}=UK
\label{Eq:PMNS}
\end{eqnarray}
with
\begin{eqnarray}
U &=&\left( \begin{array}{ccc}
  c_{12}c_{13} &  s_{12}c_{13}&  s_{13}e^{-i\delta}\\
  -c_{23}s_{12}-s_{23}c_{12}s_{13}e^{i\delta}
                                 &  c_{23}c_{12}-s_{23}s_{12}s_{13}e^{i\delta}
                                 &  s_{23}c_{13}\\
  s_{23}s_{12}-c_{23}c_{12}s_{13}e^{i\delta}
                                 &  -s_{23}c_{12}-c_{23}s_{12}s_{13}e^{i\delta}
                                 & c_{23}c_{13}\\
\end{array} \right),
\label{Eq:U_nu}
\\
K &=& {\rm diag}(e^{i\rho}, e^{i\sigma}, 1),
\label{Eq:K}
\end{eqnarray}
Where $c_{ij}=\cos\theta_{ij}$ and $s_{ij}=\sin\theta_{ij}$, the phases $\rho$ and $\sigma$ are physically relevant only if the massive neutrinos are Majorana particles.\\

The mixing angles $\theta_{23}$ and $\theta_{12}$ are much larger than $\theta_{13}$. Before the recent experiment results of T2K \cite{T2K}, MINOS \cite{MINOS} and Double Chooz \cite{DC}, the best-fit of neutrino oscillation angles are usually taken to be $\theta_{13}\approx 0^\circ$, $\theta_{23}\approx 45^\circ$ and $\theta_{12}\approx 34^\circ$. Many ansatzes are proposed to explain the smallness of $\theta_{13}$ and maximality of $\sin\theta_{23}$ \cite{Xing01}. The well-known tri-bimaximal mixing pattern is consistent with experiment data then \cite{HPS}. But Daya Bay SBL experiment made a precise measurement of $\theta_{13}$ from the $\overline{\nu}^{}_e \to \overline{\nu}^{}_e$ oscillations \cite{DB}. The $\theta_{13}$ best-fit ($\pm 1\sigma$ range) result is
\begin{eqnarray}
\sin^2 2\theta^{}_{13} = 0.092 \pm 0.016 ({\rm stat})
\pm 0.005 ({\rm syst}) \; ,
\end{eqnarray}
which is to say that $\theta^{}_{13} \neq 0^\circ$ at the $5.2\sigma$
level. The RENO collaboration confirmed the Daya Bay result soon \cite{RE}.\\

$\theta_{13}$ is not really very small. This fact makes many ansatzes face difficulties \cite{Xing01} and opens the door of possible new phenomenological applications. The possible explanation and impact of large $\theta_{13}$ have been discussed recently by many authors for example \cite {WYL}-\cite{GE}. The precise measurement of $\theta_{13}$ also has impact on the global analysis of the mixing parameters. Many groups have already performed global analyses recently \cite{FO} \cite{VA}. We pay particular attention to the result of \cite{FO}. There is an interesting indication that the mixing angle $\theta_{23}$ deviates from the maximal mixing(i.e.,$\theta_{23} < \frac{\pi}{4}$(at $\leq$ 3$\sigma$ in NH and $\leq$ 2$\sigma$ in IH). And a weak hint at the CP-violation phase $\delta \sim \pi$. But a reliable result on $\delta$ is not available without the future long-baseline neutrino oscillation experiments. \\

In the phenomenological view, the new global analysis result may change the form of the PMNS matrix significantly. If the value of $\theta_{23}$ is much smaller, the absolute value of $U_{\tau 1}$ ($|U_{\tau 1}|=|s_{12}s_{23}-c_{12}c_{23}s_{13}e^{i\delta}|$) will also decrease. This makes the Hermitian (or symmetric) ansatz of PMNS matrix seem more possible. In fact, a Hermitian(and symmetric) ansatz has been discussed by some authors \cite{HR}\cite{JS}. The phenomenological view of the Hermitian PMNS matrix also appears in a recent paper \cite{BZ}. Encouraged by the global analysis result \cite{FO}, we study the possible view that the PMNS matrix is Hermitian in the leading order. Consider the strategy that the real PMNS matrix has the following structure \cite{Xing01},
\begin{eqnarray}
U_{PMNS}=\left(U^{}_h + \Delta U\right)K \; ,
\end{eqnarray}
where $U_h$ is the Hermitian ansatz in the leading order, and $\Delta U$ is the perturbation.\\

We propose a Hermitian anzatz according to the recent global analysis data in section 2. The relation between the possible perturbation and the mass matrix of lepton is discussed in section 3. And we calculate next-to-the-leading and next-to-next-to-the-leading corrections. We find that it has a prediction consistent with the data very well. Further discussion  of the neutrino mass matrix and conclusion appears in section 4.

\section{The ansatz}

To get a proper ansatz, we turn to the experiment data \cite{FO}. We simply use the global analysis best-fit data in the standard parametrization (3). As observed by Yoni-Ben Tov and Zee \cite{BZ}, the exact Hermitian mixing matrix would predict the conservation of CP. So we expect the CP-violation is very small if the Hermitian ansatz is proper. Though there is a  hint that $\delta \sim \pi$, the Hermitian ansatz seems more probable when the  $\delta \sim 0$. If the CP-violation  originates from a small perturbation of an ansatz, it is more natural to assume $\delta \sim 0$. So we just ignore the possibility that $\delta \sim \pi$,and use the NH data \cite{FO}
\begin{eqnarray}
\sin^2 \theta_{12}=0.307 ,  \sin^2 \theta_{13}=0.0241  ,  \sin^2 \theta_{23}=0.386, \delta =0
\end{eqnarray}
Then the matrix $U_{exp}$ is
\begin{eqnarray}
U_{exp}=\left(\begin{array}{ccc}
0.822 & 0.547 & 0.155\\
-0.514 &0.599 &0.614 \\
0.243 &-0.584 &0.777
\end{array}\right)
\end{eqnarray}
The minus sign of $U_{exp,}{_{\mu 1}}$ and $U_{exp,}{_{\tau 2}}$ can be changed by making a basis transformation
\begin{eqnarray}
e_{L} \to diag \left(\begin{array}{ccc} 1, & 1, & e^{i\pi}\end{array}\right)e_{L},
\end{eqnarray}
 and
\begin{eqnarray}
\nu_{L} \to diag\left(\begin{array}{ccc} e^{i\pi},&1 & ,1 \end{array}\right)\nu_{L}
\end{eqnarray}
a phase of K would change, if the neutrinos are Majorana. Then the matrix $U_{exp}$ would look proximately Hermitian
\begin{eqnarray}
U_{exp}=\left(\begin{array}{ccc}
-0.822 & 0.547 & 0.155\\
0.514 &0.599 &0.614 \\
0.243 &0.584 &-0.777
\end{array}\right)
\end{eqnarray}
Encouraged by the data, we put forward an simple ansatz of the matrix U. The form of this matrix is the following:
\begin{eqnarray}
U^{(0)}_h=\left(\begin{array}{ccc}
-a & b & c\\
 b & b &b \\
c & b & -a
\end{array}\right)
\end{eqnarray}
This form of the mixing matrix also exhibits symmetry in the lepton sector. The general real and Hermitian  matrix is
\begin{eqnarray}
U^{(0)}_h=\left(\begin{array}{ccc}
-a & b & c\\
 b & d &e \\
c & e & f
\end{array}\right)
\end{eqnarray}
If the $Z_2$ transformation  $\nu_1 \leftrightarrow \nu_3$ and $e \leftrightarrow \tau$ keeps the mixing matrix invariant, we have  $-a=f$ and $b=e$. We just need  $d=b$ to get the form (12), it turns out that this requirement results from a certain form of the mass matrix, as will be discussed in
sect.4.
The unitary property of matrix U helps us to fix the parameters a, b and c. The numerical result is
\begin{eqnarray}
U^{(0)}_h=\left(\begin{array}{ccc}
\frac{1}{\sqrt{3}-3}& \frac{1}{\sqrt{3}} & \frac{1}{\sqrt{3}+3}\\
 \frac{1}{\sqrt{3}}  & \frac{1}{\sqrt{3}}  & \frac{1}{\sqrt{3}}  \\
\frac{1}{\sqrt{3}+3} & \frac{1}{\sqrt{3}}  &\frac{1}{\sqrt{3}-3}
\end{array}\right)
\end{eqnarray}
$ U^{(0)}_h $ has a simple and beautiful structure. Comparing with the parametrization (3), we get $\theta_{13} \approx 12.2^{\circ}$ deviating from experiment result about $3^\circ$ in this simple mixing pattern, and  the degeneration of the $\theta_{23}$ and $ \theta_{12}$, and of course no CP-violation.  According to (6), the deviation of our ansatz from the best-fit data is
\begin{eqnarray}
\Delta U=\left(\begin{array}{ccc}
-0.0333 & -0.0304 & -0.0563\\
 -0.0634 & 0.0216 &0.0366 \\
0.0317 & 0.00660 & 0.0117
\end{array}\right)
\end{eqnarray}
The magnitude of each element of $\Delta U$ is of $O(0.01)$($U_{exp}$ is of $O(0.1)$). This implies that all the mixing angles should receive the same order correction in a natural way \cite{Xing01}. The relative deviation of $\theta_{13}$ is much larger than others because of the smallness  of $\theta_{13}$. We will show in section 3 that it is easy to make a perturbation to match the experimental data.\\

\section{High order corrections}

The lepton flavor mixing matrix $U_{PMNS}$ is directly related with the diagonalization of the charged lepton mass matrix $M_l$ and neutrino mass matrix $M_{\nu}$. We take the charged lepton mass matrix to be Hermitian and consider the case that the neutrinos are Majorana (There is only the left-hand Majorana mass term). So the mass matrix can be diagonalized as
\begin{eqnarray}
V_{l}^{\dagger}M_{l}V_{l}=diag \left(\begin{array}{ccc}m_e,&m_{\mu},&m_{\tau}\end{array}\right)
\end{eqnarray}
and
\begin{eqnarray}
V_{\nu}^{T}M_{\nu}V_{\nu}=diag \left(\begin{array}{ccc}m_1,&m_2,&m_3\end{array}\right)
\end{eqnarray}
The PMNS matrix is $U_{PMNS}=V_{l}^{\dagger}V_{\nu}$. In general, it is the product of $ V_{l}^{\dagger}$ and $V_{\nu}$ that determines the mixing pattern. But the possibility that the leading effect of mixing pattern is caused by $V_{\nu}$ or $V_{l}$ with the other one as the next-leading order perturbation has been suggested and studied \cite{LP}. The structure of mass matrix is directly related with the transformation matrix. The mass spectrum of charged lepton and neutrino would give some information about the mass matrix structure. The much larger mass hierarchy of charged leptons may have some relation with the mass matrix elements hierarchy structure. And the mass hierarchy of neutrino is not so large, so it may be natural to think that the mass matrix elements are not in the hierarchy structure. One possible and simple charged lepton mass matrix structure is that it is nearly diagonalized. Then in the leading order the transformation matrix of charged lepton is close to 1. So it is the transformation matrix $V_\nu$ that determines the leading mixing pattern.\\

We start from the charged lepton mass matrix:
\begin{eqnarray}
M_l=\left(\begin{array}{ccc}
m_1 & \epsilon & \eta\\
 \epsilon ^{*} & m_2 &\kappa \\
\eta ^{*} & \kappa ^{*} & m_3
\end{array}\right)
\end{eqnarray}
Where $m_1$,$m_2$ and $m_3$ are all real, $\epsilon$, $\eta$ and $\kappa$ are complex. Considering the mass hierarchy of charged leptons, we assume the following relation between these parameters.
\begin{eqnarray}
|\epsilon|,m_1\ll m_2,m_3,
\end{eqnarray}
\begin{eqnarray}
|\eta|,|\kappa| \ll |\epsilon| ,m_1\quad and \quad
m_1 \ll |\epsilon|
\end{eqnarray}
This charged leptons mass matrix is nearly diagonalized. The leading order structure is assumed to be the following.
\begin{eqnarray}
M_l^{(0)}=\left(\begin{array}{ccc}
0 & 0 &0\\
 0  & m_2 &0 \\
0 & 0 & m_3
\end{array}\right)
\end{eqnarray}
This predicts the mass of electron be zero. This form is reasonable for the big mass hierarchy between e and $\mu$ ($\frac{m_{\mu}}{m_e} \approx 200$). The untiray transformation matrix $V_l$ is trivial, i.e. $V_l^{(0)}=1$. So the transformation matrix $V_\nu$ is just the leading mixing matrix $U_h^{(0)}$. And we assume the transformation matrix $V_\nu$ stays the same in the all order calculation. The next-leading order is just related with the structure of the charged lepon mass matrix.\\

Before we go on to consider the next to the leading order, let us compare the leading order prediction with the data. The leading order predicts that $\sin^2 \theta_{12}^{(0)}= \sin^2 \theta_{23}^{(0)} \approx 0.349$. $\theta_{23}^{(0)}$ is in the $2\sigma$ region and $\theta_{12}^{(0)}$ is just in the $3\sigma$ region. $\theta_{13}^{(0)}$ is even worse, since $\sin^2 \theta_{13}^{(0)} \approx 0.0446$, which is out of the $3\sigma$ allowed region of the data, so the next-leading order calculation is necessary.\\

Consider (19), the next-leading order mass matrix of charged lepton is \cite{Xing02}
\begin{eqnarray}
M_l^{(1)}=\left(\begin{array}{ccc}
0 & \epsilon &0\\
 \epsilon^{*}  & m_2 &0 \\
0 & 0 & m_3
\end{array}\right)
\end{eqnarray}
With $\epsilon =|\epsilon|e^{i\lambda}$.
This matrix deviates from the diagonalized matrix by off-diagonal element $\epsilon$, but it is also easy to be diagonalized by the transformation matrix
\begin{eqnarray}
V_l^{(1)}=\left(\begin{array}{ccc}
\cos \theta & \sin \theta e^{i\phi}& 0\\
   -sin \theta e^{-i\phi} & \cos \theta &0 \\
0 & 0 & 1
\end{array}\right)
\end{eqnarray}
Where $\tan \theta =\sqrt {\frac {m_e}{m_\mu}} \approx 0.07$, $m_3=m_\tau$, $m_2=m_\mu-m_e$, $|\epsilon|=\sqrt {m_em_\mu}$ and $\phi =\lambda$. Using the relation $U^{(1)}_h=V_l^{(1)\dagger}V_\nu=V_l^{(1)\dagger}U_h^{(0)}$, we get the next-leading order mixing matrix:
\begin{eqnarray}
U^{(1)}_h=\left(\begin{array}{ccc}
\frac{\cos \theta}{\sqrt 3 -3}-\frac{\sin \theta e^{i\lambda}}{\sqrt 3} & \frac{\cos \theta}{\sqrt 3}-\frac{\sin \theta e^{i\lambda}}{\sqrt 3}  &\frac{\cos \theta}{\sqrt 3 +3}-\frac{\sin \theta e^{i\lambda}}{\sqrt 3} \\
 \frac{\cos \theta}{\sqrt 3 }+\frac{\sin \theta e^{-i\lambda}}{\sqrt 3 -3}   & \frac{\cos \theta}{\sqrt 3 }+\frac{\sin \theta e^{-i\lambda}}{\sqrt 3}  &\frac{\cos \theta}{\sqrt 3}+\frac{\sin \theta e^{-i\lambda}}{\sqrt 3 +3}  \\
\frac {1}{\sqrt 3 +3} & \frac {1}{\sqrt 3} & \frac {1}{\sqrt 3-3}
\end{array}\right)
\end{eqnarray}
There is only one free parameter $\lambda$ here. And the possible CP-violation phase is also included in the next-leading order. Now after rephasing (23) and comparing with the standard parametrization (3) it is straightforward to write down the mixing parameters after correction,
\begin{eqnarray}
\sin ^2 \theta_{13}^{(1)}\simeq (\frac {1}{\sqrt 3 +3})^2-\frac{2}{3+3\sqrt 3}\sqrt {\frac {m_e}{m_\mu}} \cos \lambda,
\end{eqnarray}
\begin{eqnarray}
\sin ^2 \theta_{23}^{(1)} \simeq \frac{2+2(\sqrt 3 -1) \sqrt {\frac {m_e}{m_\mu}}\cos \lambda}{4+\sqrt 3+2(\sqrt 3 -1) \sqrt {\frac {m_e}{m_\mu}} \cos \lambda},
\end{eqnarray}
\begin{eqnarray} 	
\sin ^2 \theta_{12}^{(1)} \simeq \frac{2- 4\sqrt {\frac {m_e}{m_\mu}}\cos \lambda}{4+\sqrt 3+2(\sqrt 3 -1) \sqrt {\frac {m_e}{m_\mu}}\cos \lambda},
\end{eqnarray}
\begin{eqnarray} 	
\tan \delta^{(1)} \simeq \frac{-(\sqrt 3 +1)\sqrt {\frac {m_e}{m_\mu}}\sin \lambda}{1-(1+\sqrt 3)\sqrt {\frac {m_e}{m_\mu}}\cos \lambda},
\end{eqnarray}
Here we only keep the order $O(\sqrt {\frac {m_e}{m_\mu})}$. If $\lambda$ is in the first and fourth quadrant, the degeneration of $\theta_{23}$ and $\theta_{12}$ disappers with $\theta_{23}$ becoming larger and $\theta_{12}$ becoming smaller, which is in keep with the experiment data. The next-leading order correction to $\sin ^2 \theta_{13}^{(0)}$ is about -0.02, which also can fit the data much better. More precisely, $\theta_{13}^{(1)}$ can lie in the $2\sigma$ range if
$0\leq \lambda \leq 0.413$ or $5.870 \leq \lambda \leq 2\pi$,  $\theta_{13}^{(1)}$ can lie in the $3\sigma$ range if
$0\leq \lambda \leq 0.687$ or $5.596 \leq \lambda \leq 2\pi$ (see figure 1). $\theta_{23}^{(1)}$ can be in the $2\sigma$ range in a large area of the parameter space of $\lambda$. $\theta_{23}^{(1)}$ fits the data very well in the leading order, when $\lambda$ is small or near $2\pi$, $\theta_{12}^{(1)}$ would lie in the $1\sigma$ region. The best-fit point can also be included. There are two possible CP-violation phases, one is near zero, another is near $\pi$. As we have assumed, $\delta$ should be near zero in the natural way, $\lambda$ is constrained in the fourth quadrant. The rephasing-invariant parameter $\cal J$ $\approx \frac{\sqrt 3}{18}\sqrt {\frac {m_e}{m_\mu}}\sin \lambda  \sim O(0.001)$. \\

As we can see, the next-leading order correction can fit the data. But it is still not very satisfied, let us consider the next-to-next-to-the-leading order correction now. The mass matrix of charged lepton would be of the following structure
 \begin{eqnarray}
M_l^{(2)}=\left(\begin{array}{ccc}
m_1 & \epsilon &0\\
 \epsilon^{*}  & m_2 &0 \\
0 & 0 & m_3
\end{array}\right)
\end{eqnarray}
It is also very easy to diagonalize by a similar matrix
\begin{eqnarray}
V_l^{(2)}=\left(\begin{array}{ccc}
\cos \zeta & \sin \zeta e^{i\rho}& 0\\
   -sin \zeta e^{-i\rho} & \cos \zeta &0 \\
0 & 0 & 1
\end{array}\right)
\end{eqnarray}
Where $\tan \zeta =\sqrt {\frac{m_e+m_1}{m_\mu}} $, $m_3=m_\tau$, $m_2+m_1=m_\mu-m_e$ and $\rho =\lambda$. Besides the parameter $\lambda$, a new parameter appears, i.e. $m_1$. We define $m_1 \equiv (C^2-1)m_e$(with $C>1$), where C is real and positive. Then $\tan \zeta =C\sqrt {\frac{m_e}{m_\mu}} $. As we have assume $m_1 \ll |\epsilon| $ and know $|\epsilon|$ is of O($\sqrt {m_\mu m_e}$)$\sim O(10)m_e$, $m_1$ should be order $m_e$, i.e. C is of O(1). It is straightforward to calculate the next-to-next-to-the-leading order correction as the same way above
\begin{eqnarray}
\sin ^2 \theta_{13}^{(2)}\simeq (\frac{1}{\sqrt 3 +3})^2-\frac{2}{3+3\sqrt 3}\sqrt {\frac {m_e}{m_\mu}}C \cos \lambda,
\end{eqnarray}
\begin{eqnarray}
\sin ^2 \theta_{23}^{(2)} \simeq \frac{2+2(\sqrt 3 -1) \sqrt {\frac {m_e}{m_\mu}}C\cos \lambda}{4+\sqrt 3+2(\sqrt 3 -1) \sqrt {\frac {m_e}{m_\mu}}C\cos \lambda},
\end{eqnarray}
\begin{eqnarray} 	
\sin ^2 \theta_{12}^{(2)} \simeq \frac{2- 4\sqrt {\frac {m_e}{m_\mu}}C\cos \lambda}{4+\sqrt 3+2(\sqrt 3 -1) \sqrt {\frac {m_e}{m_\mu}}C\cos \lambda},
\end{eqnarray}
\begin{eqnarray} 	
\tan \delta^{(2)} \simeq \frac{-(\sqrt 3 +1)\sqrt {\frac {m_e}{m_\mu}}C\sin \lambda}{1-(1+\sqrt 3)\sqrt {\frac {m_e}{m_\mu}}C\cos \lambda},
\end{eqnarray}
The new parameter C is a linear decreasing function of $\theta_{13}$, so $\theta_{13}$ is sensitive to the value of C. It is possible to make $\sin^2 \theta_{13}^{(2)}$  lie in the $1\sigma$ region. $\sin ^2 \theta_{12}^{(2)}$ can still be in the $1\sigma$ region in a large interval of C and $\lambda$. $\sin ^2 \theta_{23}^{(2)}$ is an increasing function of C, if $\lambda$ is fixed in the first or fourth quadrant. But $\sin ^2 \theta_{23}^{(2)}$ increases very slowly, so a large value of C is needed to get the $1\sigma$ region data. On the other hand, too large C would make  $\sin ^2 \theta_{12}^{(2)}$ and $\sin^2 \theta_{13}^{(2)}$ decrease too much. Though $\lambda$ could be adjusted to reduce $\sin ^2 \theta_{12}^{(2)}$, the parameter space of $\lambda$ and C is expected to be narrow. We expect the higher order correction will affect $\sin ^2 \theta_{23}^{(2)}$. But it is easy to find C and $\lambda$ to keep  $\sin^2 \theta_{13}^{(2)}$ and $\sin^2 \theta_{12}^{(2)}$ both in the $1\sigma$ region and $\sin^2 \theta_{23}^{(2)}$ in the $2\sigma$ region of data. The CP-violation phase $\delta^{(2)}$ is still near zero with a deviation about $10^{\circ}$. Future measurement of $\delta$ will test our theoretical consideration.\\

We do not consider higher order corrections here.

\section{The neutrino mass matrix}
In this last section, we make some comments on the mass matrix of neutrinos.  The hierarchy structure of charged lepton mass matrix elements give the nearly diagonalized matrix in the leading order. The transformation matrix $V_\nu$ is just the product of leading order mixing matrix $U_h^{(0)}$ and K, i.e. $V_\nu$ = $U_h^{(0)}K$ . Using (16), we can get the mass matrix of neutrinos as the following
\begin{eqnarray}
M_{\nu}=V_{\nu}diag \left(\begin{array}{ccc}m_1,&m_2,&m_3\end{array}\right)V_{\nu}^{T}
\end{eqnarray}
Taking $V_\nu$=$U_h^{(0)}K$ into (34), we get $M_\nu$
\begin{eqnarray}
M_\nu=\left(\begin{array}{ccc}
x+y & z & y+z\\
 z & x &2y+z \\
y+z & 2y+z & x-y
\end{array}\right)
\end{eqnarray}
With
\begin{eqnarray}
x=\frac{1}{3}(\hat{m_1}+\hat{m_2}+m_3)
\end{eqnarray}
\begin{eqnarray}
y=\frac{1}{\sqrt 3}(\hat{m_1}-m_3)
\end{eqnarray}
\begin{eqnarray}
z=\frac{1}{3}(\hat{m_2}+\frac {1}{\sqrt 3 +1}m_3-\frac {1}{\sqrt 3 -1}\hat{m_1})
\end{eqnarray}
where $\hat{m_1}=m_1e^{2i\rho}, \hat{m_2}=m_2e^{2i\sigma}$.
From the oscillation experiment and the cosmology observation, we know the mass spectrum is $m_1 \approx m_2$ ,$m_1(m_2) \ll m_3$ or $m_3 \ll m_1(m_2)$. So  elements of neutrino mass matrix do not have the hierarchy structure as the charged lepton mass matrix. But (35) has an interesting property. The sums of every column and every row are all equal, $\sum_{i=1,2,3} M_{\nu,}{_{\lambda i}}=x+2y+2z$ ($\lambda =e,\mu,\tau$) and $\sum_{\lambda=e,\mu,\tau} M_{\nu,}{_{\lambda i}}=x+2y+2z$ (i =1,2,3). This property also appears in some $A_4$ models \cite{GFMS}\cite{GM}, but mass matrix (36) show less symmetry than the $A_4$ models mass matrix. \\

The logic can also be reversed. If the mass matrix has the structure as (35), we can diagonalize it and get the mixing ansatz (13). We expect some new symmetry or underlying mechanism would give the mass matrix structure (18) and (36). Then we can know more about the origin of the mass matrix structure, which may be related with the high-scale physics. We also notice the different mass form of charged lepton and neutrino. But it is not unnatural that the mass matrix origins of charged leptons and neutrinos may be due to different mechanisms. \\

We show in this paper that a Hermitian ansatz can be made to fit the experimental data very well. The mixing ansatz that we propose in section 2 is just a phenomenological consideration, its theoretical mechanism is unclear yet. The future precise measurements of the mixing parameters, especially  of the CP-violation phase $\delta$ in the LBL experiments, will provide more information on the mixing pattern and constrain all the parameters. We expect $\delta$ be near zero, and then the Hermitian ansatz can be an important hint at the underlying physical theory.


\begin{figure}[htb]

\centering

\includegraphics[width=12cm]{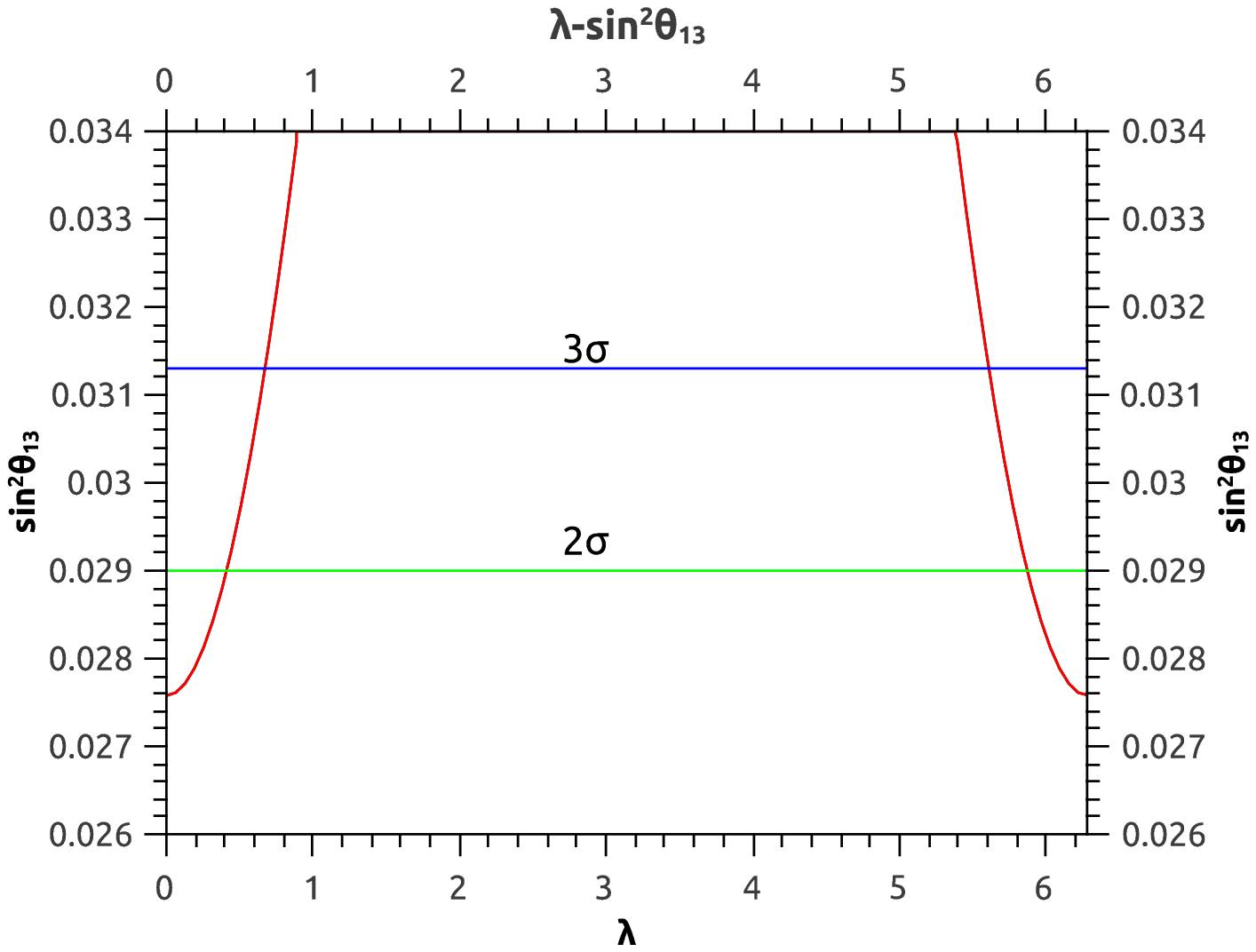}

\caption{$\sin^2 \theta_{13}$ as a function of $\lambda$ in the next-to-the-leading order } \label{fig:graph}

\end{figure}
\begin{figure}[htb]

\centering

\includegraphics[width=12cm]{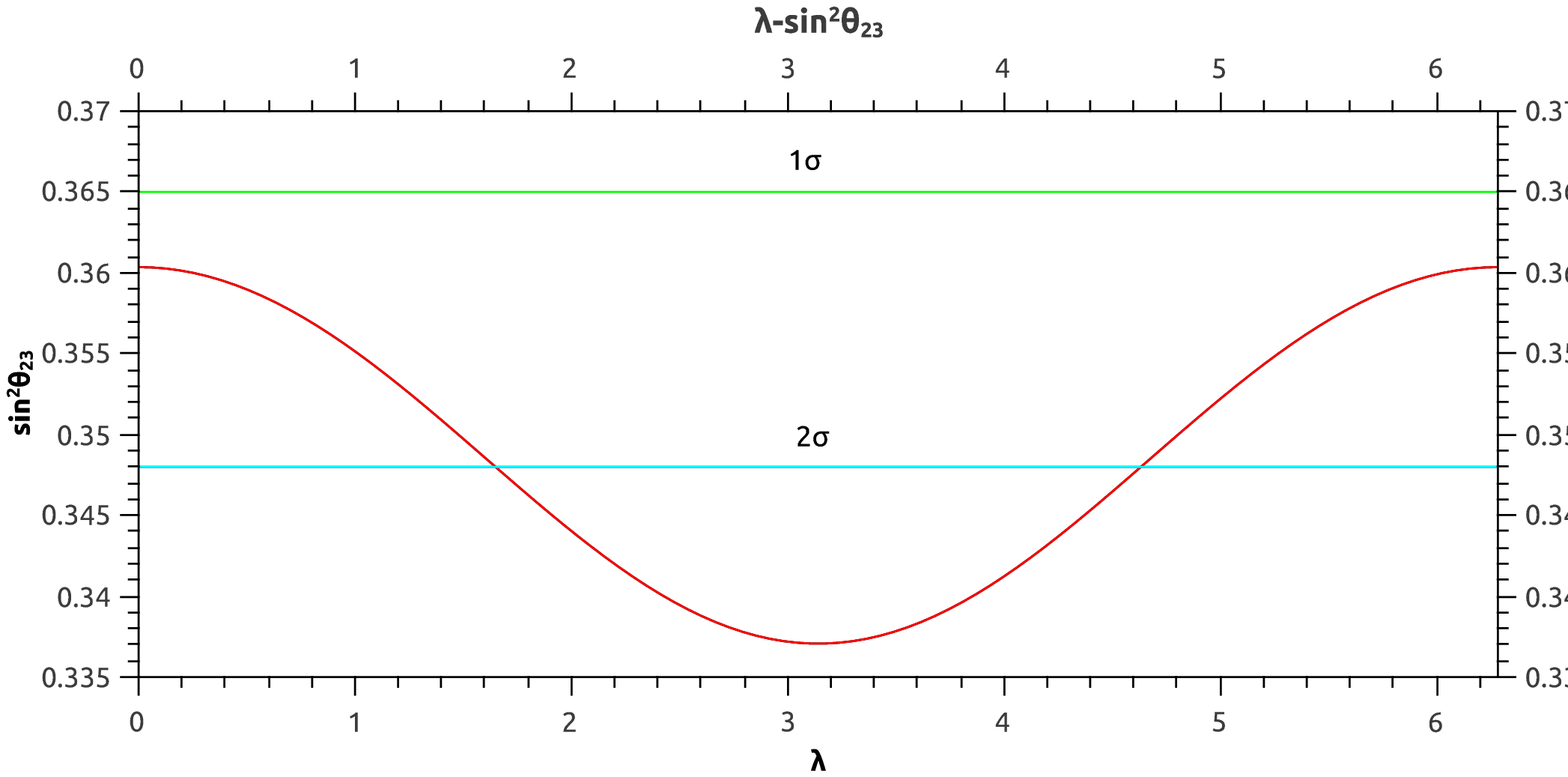}

\caption{$\sin^2 \theta_{23}$ as a function of $\lambda$ in the next-to-the-leading order } \label{fig:graph}

\end{figure}

\begin{figure}[htb]

\centering

\includegraphics[width=12cm]{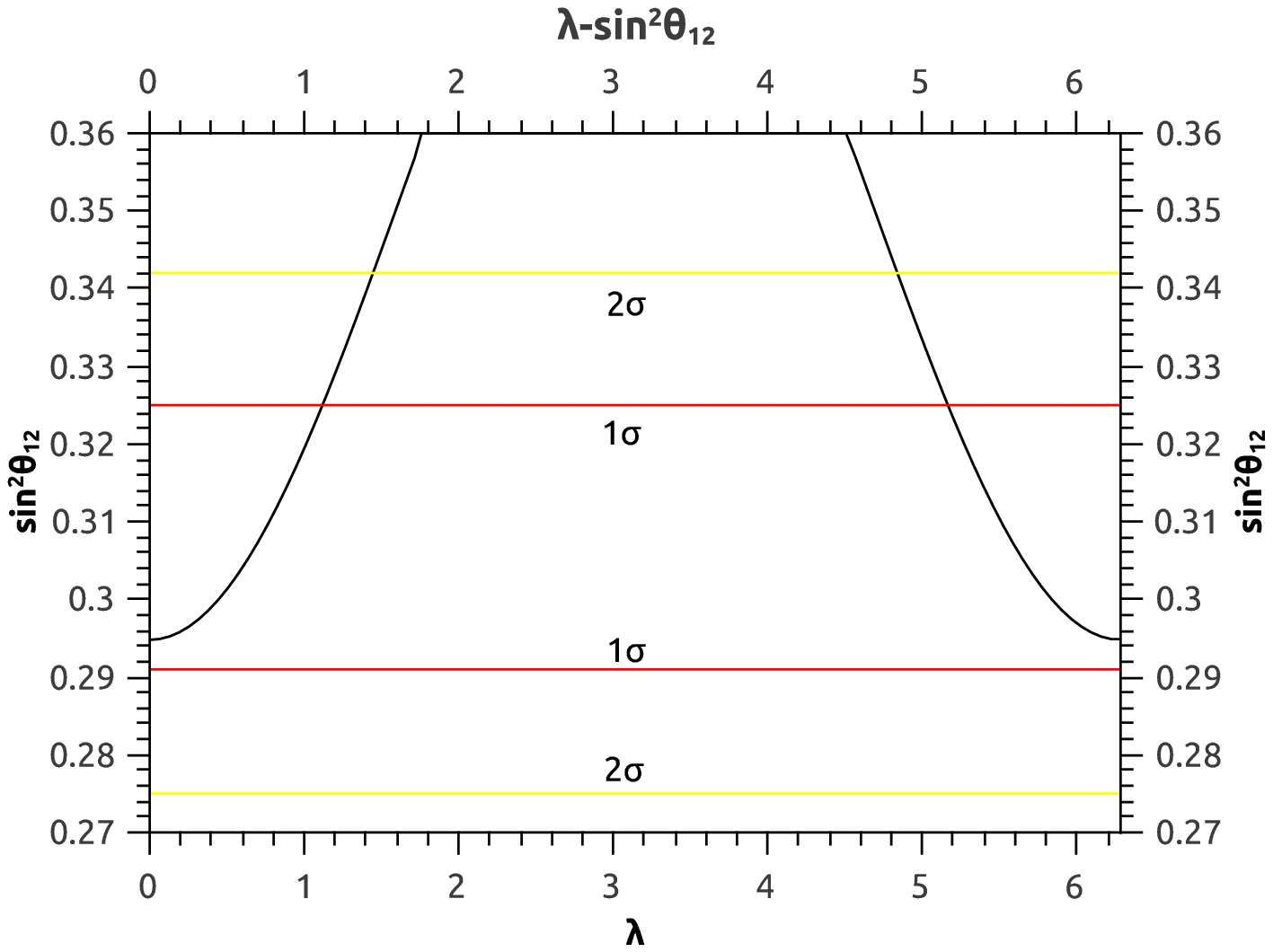}

\caption{$\sin^2 \theta_{12}$ as a function of $\lambda$ in the next-to-the-leading order } \label{fig:graph}

\end{figure}
\begin{figure}[htb]

\centering

\includegraphics[width=12cm]{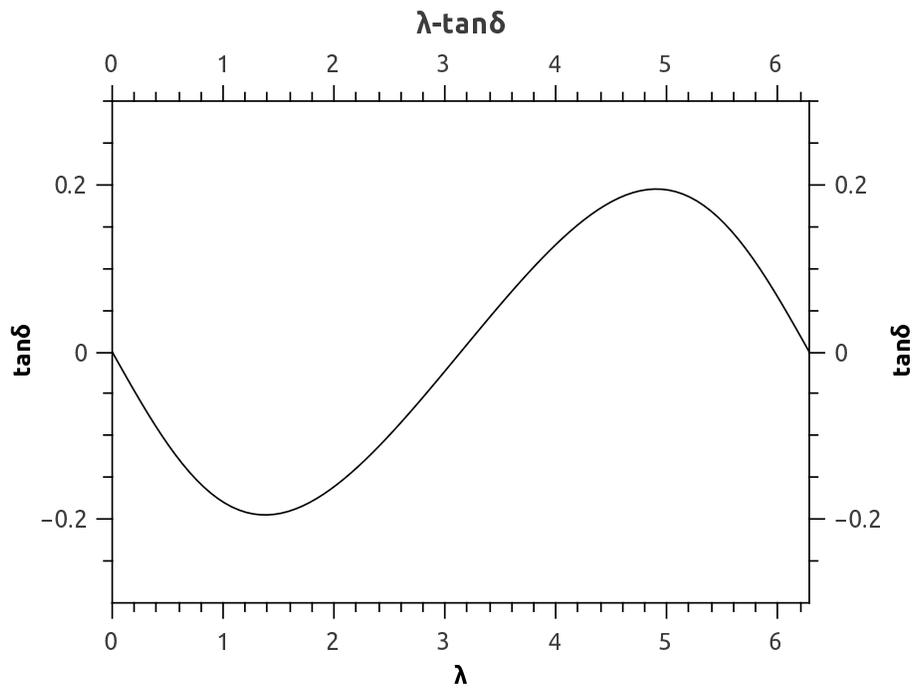}

\caption{$\tan \delta$ as a function of $\lambda$ in the next-to-the-leading order } \label{fig:graph}
\end{figure}

\end{document}